\documentclass[iop]{emulateapj}
\pdfoutput=1

\begin{document}
\title{The Effect of the Pre-Detonation Stellar Internal Velocity Profile on the Nucleosynthetic Yields in Type Ia Supernova}
\author{Yeunjin Kim\altaffilmark{1}, G. C. Jordan IV\altaffilmark{1,2}, Carlo Graziani\altaffilmark{1,2}, B. 
S. Meyer\altaffilmark{3}, D. Q. Lamb\altaffilmark{1,2,4}, \\J. W. Truran\altaffilmark{1,2,5,6}}

%Specify affiliation information
\altaffiltext{1}{Astronomy Department, The University of Chicago, Chicago, IL 60637}
\altaffiltext{2}{Flash Center Computational Science, The University of Chicago, Chicago, IL 60637}
\altaffiltext{3}{Physics and Astronomy Department, Clemson University, Clemson, SC 29634}
\altaffiltext{4}{Enrico Fermi Institute, The University of Chicago, Chicago, IL 60637} 
\altaffiltext{5}{Joint Institute for Nuclear Astrophysics, University of Chicago, Chicago, IL 60637}
\altaffiltext{6}{Argonne National Laboratory, Argonne, IL, 60439}

\begin{abstract}
A common model of the explosion mechanism of Type Ia supernovae is based on a delayed 
detonation of a white dwarf. A variety of models differ primarily in the method by which the 
deflagration leads to a detonation. A common feature of the models, however, is that all of them 
involve the propagation of the detonation through a white dwarf that is either expanding or 
contracting, where the stellar internal velocity profile depends on both time and space. In this work, we 
investigate the effects of the pre-detonation stellar internal velocity profile and the post-detonation 
velocity of expansion on the production of $\alpha$-particle nuclei, including $^{56}$Ni, which are the 
primary nuclei produced by the detonation wave. We perform one-dimensional hydrodynamic 
simulations of the explosion phase of the white dwarf for center and off-center detonations with five 
different stellar velocity profiles at the onset of the detonation. In order to follow the complex flows and 
to calculate the nucleosynthetic yields, 
approximately 10,000 tracer particles were added to every simulation. We observe two distinct 
post-detonation expansion phases: rarefaction and bulk expansion. Almost all the burning to 
$^{56}$Ni occurs only in the rarefaction phase, and its expansion time scale is influenced 
by pre-existing flow structure in the star, in particular by the pre-detonation stellar velocity profile. We 
find that the mass fractions of the $\alpha$-particle nuclei, including $^{56}$Ni, are tight functions of 
the empirical physical parameter $\rho_{up}$/$v_{down}$, 
where  $\rho_{up}$ is the mass density immediately upstream of the detonation wave front and 
$v_{down} $ is the velocity of the flow immediately downstream of the detonation wave front. We also 
find that $v_{down}$ 
depends on the pre-detonation flow velocity. We conclude that the properties of the pre-existing flow, 
in particular the internal stellar velocity profile, influence the final isotopic 
composition of burned matter produced by the detonation.     
\end{abstract}
\keywords{nuclear reactions, nucleosynthesis, abundances; supernovae: general; white dwarfs}

\section{Introduction}

Type Ia supernovae (SNe Ia) result from the explosions of carbon-oxygen (CO) white dwarfs 
\citep{Fow60}. Their extremely high and very similar peak luminosities
have made SNe Ia important standard candles for measuring distances in the Universe, and therefore 
estimating cosmological parameters
critical for our understanding of its evolution \citep{Per98,Sch98,Rie01}. %(Ivo) 
They also play a critical role in nucleosynthesis, as contributors to the abundances of both 
iron-peak (titanium to zinc) nuclei and intermediate mass (silicon to calcium) elements. 

Despite the widespread use of and interest in SNe Ia, their explosion mechanism is not yet fully 
understood, and their progenitor systems have not been unambiguously identified. We 
have chosen to explore the single-degenerate scenario of SNe Ia, which involves the growth of a 
Carbon-Oxygen white dwarf toward the Chandrasekhar limit in a binary system and its ultimate 
thermonuclear ignition 
and disruption. The initial phases of the explosion involve a subsonic thermonuclear flame 
(deflagration) which gives rise to the expansion of the white dwarf prior to the initiation of a 
supersonic burning front (detonation wave) \citep{Kho91, Hof95, Nie97}. The density of the white 
dwarf decreases as a consequence of its expansion. Thus, the detonation burns material at 
lower densities, producing higher abundances of intermediate mass elements \citep{Hil00}. 
According to observations, approximately 0.4 solar masses of intermediate mass 
elements are synthesized during the explosion \citep{Whe90}. A variety of models
involve the scenario of a deflagration phase followed by a detonation phase, such as the 
deflagration-to-detonation transition (DDT)\citep{Kho91,Gam04}, the pulsating reverse detonation 
(PRD) \citep{Bra06,Bar09,Bra09}, or the gravitationally confined detonation (GCD) \citep{Ple04, Tow07, Jor08, Mea09, Jor12}. These models differ primarily in the method by which the deflagration 
phase leads to a detonation wave. A common feature of the models, however, is that all of them 
involve the 
propagation of the detonation wave through a white dwarf that is either expanding or contracting, 
where the stellar velocity profile depends on both time and space. The detonation wave thus 
propagates into material that is not at rest. 

In this paper, we explore the post-detonation expansion of the white dwarf and its consequences for 
the production of $\alpha$-particle nuclei, including intermediate-mass nuclei and $^{56}$Ni, using 
central and off-center detonation models with different velocity profiles at 
detonation\footnote{A related work is the study by \citet{Gar07} of the nucleosynthetic yield produced by a subsonic deflagration wave in the low-density outer layers of an expanding white dwarf.}. 
We describe the details of the numerical methods used in our simulations  
in $\S  \ref{Numerical Methods}$. We performed a total of nine simulations; we give the 
setups for these simulations in $\S \ref{Model Setup}$. We present our results in $\S \ref{Results}$,  
focusing primarily on the central explosion models.  Our results include the detonation structure for 
non-static stars ($\S \ref{Detonation Structure for Non-Steady 
Stars}$), the expansion of the flow at different time scales behind the detonation 
and the nucleosynthesis using the thermal properties of the flow ($\S  \ref{The Effect of Expansion 
Time 
Scales on Nucleosynthesis}$), and an empirical relation between certain properties of the detonation 
wave and the final nucleosynthetic yields of $\alpha$-particle nuclei, including $^{56}$Ni, in the 
density range of  
$5 \times 10^{6}  \sim 2.5 \times 10^{7} g \ cm^{-3}$ ($\S \ref{An Empirical Relation}
$). We present the results for the off-center detonation models 
in $\S \ref{Off-Center Detonation}$. We conclude with a 
summary of our findings and of their implications for our understanding of the consequences of the 
detonation under the conditions appropriate for a pulsating white dwarf environment ($\S  
\ref{Conclusion}$).
    
\section{Numerical Methods}
\label{Numerical Methods}
\subsection{FLASH}

We performed one-dimensional spherically-symmetric simulations using the FLASH code 
\citep{Fry00}. FLASH is  a modular, block-structured Eulerian code with adaptive mesh refinement 
capabilities that solves Euler's equations for compressible gas dynamics with a directionally split 
piecewise parabolic method \citep{Col84}. PPM is particularly well-suited to flows 
involving discontinuities, such as detonation waves, and allows handling of non-ideal 
equations of state (EOS) \citep{Col85}. The implementation of the EOS uses a tabular 
Helmholtz free energy EOS appropriate for stellar conditions encountered here including 
contributions from blackbody radiation, ions, and electrons of an arbitrary degree of degeneracy
 \citep{Tim00}. 
 
For the energetics required to treat detonation waves, we track three distinct burning stages: (1) the 
burning of carbon to O, Ne, Na, Mg, and $\alpha$-particles; (2) the subsequent burning of oxygen to a 
Quasi-Static Equilibrium (QSE) state comprised predominantly of Si group 
elements; and (3) if the temperature is high enough, the evolution of the system to Nuclear Statistical 
Equilibrium (NSE), which at the low entropies found in Chandrasekhar mass white dwarf, is primarily of comprised iron-group elements. The energy 
released due to carbon burning is explicitly calculated using the carbon fusion rate of \citet{Cau88}, 
while the energy released during Oxygen and Si-group burning are 
controlled by the timescales for reaching the QSE state and NSE respectively. 
Additional details concerning the energetics are presented in \citet{Cal07}, \citet{Tow07}, 
\citet{Mea09}, and Seitenzahl et al. (2009). 

For the hydrodynamics, we use the the conservation equations for the reacting gas flow together with 
two source terms: gravity and nuclear reactions. A detonation wave is a form of combustion 
in which a shock wave propagates into the fluid, heating it to high temperatures and triggering nuclear 
reactions. The energy that is released supports the detonation wave. 
Since the detonation wave propagates supersonically, molecular diffusion, thermal 
conduction, and viscous effects are negligible. Therefore, the simplified set of the Euler equations 
for the reacting gas flow is:

\begin{equation}
\frac{\partial \rho }{\partial t} \ + \ \mathbf{\triangledown} \cdot (\rho \mathbf{v}) \ = \ 0
\end{equation}
\begin{equation}
\frac{\partial \rho \mathbf{v}}{\partial t} \ + \ \mathbf{\triangledown} \cdot (\rho \mathbf{vv}) \ +
\mathbf{\triangledown} P= \ \rho \mathbf{g}
\end{equation}
\begin{equation}
\frac{\partial \rho E}{\partial t} \ + \ \mathbf{\triangledown} \cdot [(\rho E \ + P)\mathbf{v}] = \ \rho 
\mathbf{v\cdot g} + \sum_{l}\rho q_{l}\dot{\Phi _{l}}
\end{equation}
\begin{equation}
\frac{\partial \rho \Phi_{l}}{\partial t}\ +\ \mathbf{\triangledown} \cdot (\rho\Phi_{l}\mathbf{v})\ = \ \rho
\dot{\Phi _{l}},
\end{equation}
where $\rho$ is the fluid density, $\mathbf{v}$ is the fluid velocity, $P$ is the pressure, $E$ is the 
sum of the internal energy $\epsilon $ and kinetic energy per unit mass with $E \  = \ \epsilon \ + \ 
\frac{\mathbf{v}^2}{2}$, $\mathbf{g}$ is the gravitational acceleration, $\Phi_{l}\ (l = 1, 2, 3)$ is a 
reaction progress variable that tracks the three burning stages (C-, O-, and Si-burning), 
$\dot{\Phi _{l}}$ is the reaction rate, and \textit{$q_{l}$} is the corresponding 
mass-specific energy release, (see \citet{Cal07}; \citet{Tow07}; \citet{Sei09} for further details).
Nuclear burning is suppressed within the shock to prevent instabilities. We performed the simulations 
using a one-dimensional, spherically symmetric computational
domain with radius 22,016 km.  We employed twelve levels of refinement, which
gave a maximum spatial resolution of 0.5 km during the simulations of the detonation 
phase.  

\subsection{Libnucnet}
\label{Libnucnet}
The network code used for the post processing is a version of the network code based on 
libnucnet, a library of C codes for storing and managing nuclear reaction networks \citep{Mey07}. 
The network contains 230 nuclear species ranging from neutrons, protons, and $\alpha$-particles, 
giving reliable results for the evolving abundances of the most important species, including the 
intermediate-mass elements (IME)  and iron-peak elements. The nuclear and reaction data for 
the calculations were taken from the JINA reaclib database \citep{Cyb10}. 
In this network, electron-capture rates were not included since at densities $\sim 10^{7} \ g\ cm^{-3}$, 
these reactions occur much more 
slowly than the detonation-induced explosion of the white dwarf. 

\section{Model Setup}
\label{Model Setup}
We have computed nine one-dimensional detonation models. Although supernovae are multi-dimensional phenomena, the one-dimensional treatment greatly simplifies the problem. We 
neglect the multi-dimensional effects such as a multicellular structure \citep{Gam99,Tim00}. However, 
the simplicity of the one-dimensional models allows us to draw general conclusions regarding the 
subject of primary interest to us: the nucleosynthetic yields produced by a detonation wave 
propagating across a star that is not static, but is instead either expanding or 
contracting. While our main focus is central-detonation models, we also explore the 
case in which the detonation occurs off-center.

\begin{deluxetable*}{c c c c c c}
\tablewidth{0pt}
\tablecaption{Physical Characteristics of the White Dwarf at the Time of Detonation.}
\tablehead{
\colhead{Physical Characteristics} & \colhead{$\mathbf{LCV}$\footnotemark[1]} & \colhead{$\mathbf{HCV}$\footnotemark[1]} & \colhead{$\mathbf{NOV}$} & \colhead{$\mathbf{HEV}$\footnotemark[1]} & 
\colhead{$\mathbf{LEV}$\footnotemark[1]}}
\startdata
Velocity-Radius Ratio [$sec^{-1}$] & -1/8 & -1/4 & 0 & 1/4 & 1/8\\
Central Density [$g/cm^{3}$] & $2.7 \times 10^7$ & $3.0 \times 10^7$ & $2.5 \times 10^7$ & $3.0 
\times 10^7$ & $2.7 \times 10^7$\\
\enddata
\label{init_table}
\footnotetext[1]{We investigate both central and off-center detonations for these initial profiles.}
\end{deluxetable*}

Each initial stellar model is a cold ($T = 3 \times 10^{7}K$), isothermal white dwarf (WD) in hydrostatic 
equilibrium with a central density of $2.2\times 10^{9}\ g/cm^{^{3}}$ and a radius of approximately 
2000 km. Its composition consists of equal parts of carbon and oxygen by mass 
throughout the star. 

\begin{figure*}
\centering
\includegraphics[width=8cm, scale=1.0, angle=90]{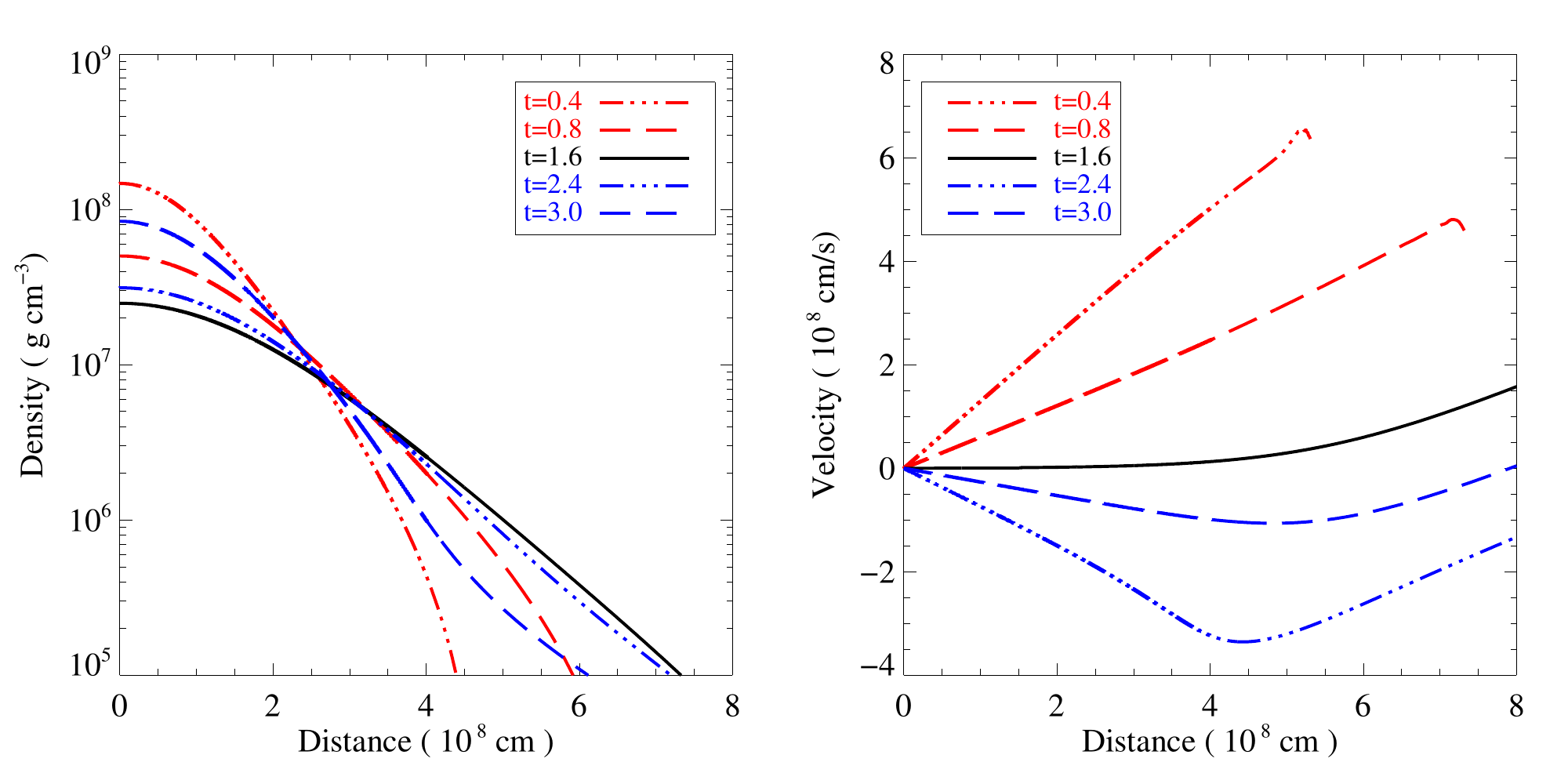}
\caption{Density profile (left panel) and velocity profile (right panel) at five different times ($t$ = 0.4, 
0.8, 1.6, 2.4, and 3.0 seconds) during the expansion phase (red), at maximum expansion (black), and 
the contraction phase (blue).  }
\label{tem_evo}
\end{figure*}

\begin{figure*}
\centering
\includegraphics[width=8cm, scale=1.0, angle=90]{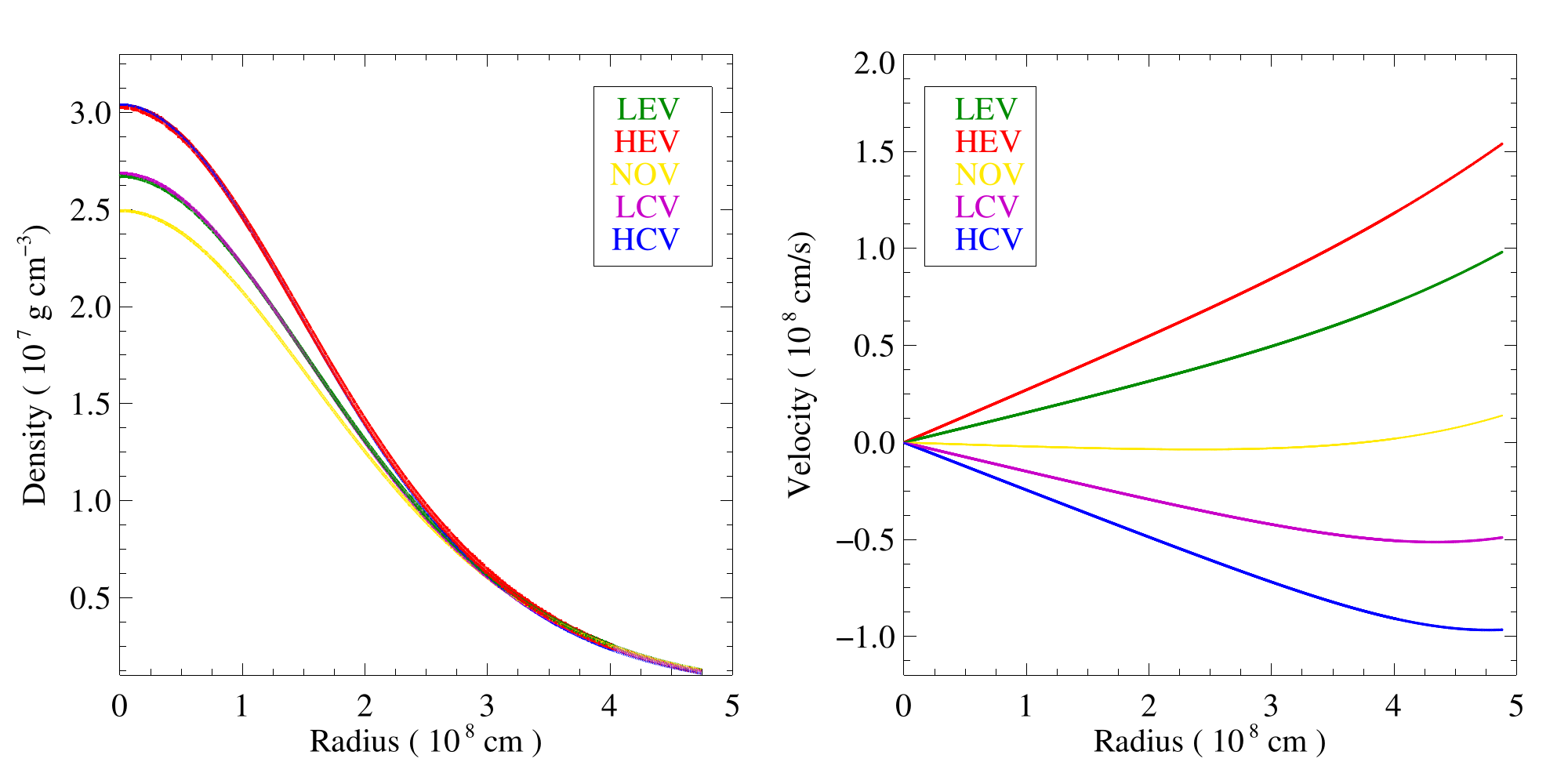}
\caption{Density profile (left panel) and velocity profile (right panel) of five one-dimensional models at 
the time of detonation. The 
relative ratio (sec$^{-1}$) between velocity and radius is approximately 1/8 ($\mathbf{LEV}$), 1/4 ($
\mathbf{HEV}$), -1/4 ($\mathbf{HCV}$), and -1/8 ($\mathbf{LCV}$). The velocity in the model $
\mathbf{NOV}$ is close to zero, and was chosen to represent the static case. In the left panel, the $
\mathbf{HEV}$ and $\mathbf{HCV}$ models have similar density profiles, while the $
\mathbf{LEV}$ and $\mathbf{LCV}$ models have similar density profiles.      }
\label{init_cond}
\end{figure*}

In the whole-star simulations of SNe Ia models that involve both deflagration and detonation phases, 
the deflagration phase leads to expansion of the WD prior to 
detonation.  In the present one-dimensional models, we artificially expand the star out of hydrostatic 
equilibrium by exciting its lowest vibrational mode with a velocity profile that varies linearly with the 
stellar radius. The 
slope of the velocity profile is such that the resulting kinetic energy is 30\% of the WD binding 
energy. This causes the star to first expand and then contract. Figure~\ref{tem_evo} shows the 
temporal evolution of the density and velocity profiles over the time during which the star expands and 
contracts. We initiated a detonation by setting the 
temperature in selected computational cells to $5.0\times 10^{9}\ K$. We detonated five WD models at 
the center of the star, while we detonated four models at $2.0\times 10^{8}cm$ slightly off-center. The 
set of central and off-center ignition models share the same initial stellar profiles and differ only by the 
ignition location. In total, we studied five initial density and velocity profiles at the time of the 
ignition; these are shown in Figure~\ref{init_cond}. The velocity profiles are nearly linear 
with radius for densities higher than $5.0\times 10^{6}\ g/cm^{^{3}}$. We were therefore able to 
choose two expanding white dwarf models in which the ratio between the
initial velocity and radius profiles is approximately 1/4 ($\mathbf{HEV}$) and 1/8 ($\mathbf{LEV}$) in 
units of sec$^{-1}$. For the contracting models, we choose two models in which the ratio between 
the initial velocity and radius profiles is approximately -1/4 ($\mathbf{HCV}$) and -1/8 ($\mathbf{LCV}
$) in units of sec$^{-1}$. Finally, we choose an initial profile ($\mathbf{NOV}$) whose velocity profile is 
approximately 0.0 sec$^{-1}$ to represent the static case. For simplicity, we denote these models by 
the ratio of the initial velocity and radius profiles throughout this paper. Unless explicitly stated 
otherwise, the data used are from the central detonation models. Table~\ref{init_table} summarizes 
the physical characteristics of the models we study.
We adopt the following nomenclature for the models we study: the first letter indicates whether the 
velocity is relatively $\mathbf{L}$ow or $\mathbf{H}$igh, the second letter indicates whether the star is  
$\mathbf{C}$ontracting or $\mathbf{E}$xpanding mode, while the third letter stands for $\mathbf{V}
$elocity.  The letters ``NOV" stand for $\mathbf{NO\ V}$elocity. 

Every simulation included Lagrangian tracer particles distributed by mass that were 
passively advected with the fluid during the course of the simulation. The temperature 
and density histories of the tracer particles were calculated by interpolating the corresponding 
quantities using the underlying Eulerian grid. 
Each tracer particle represents a fluid parcel, all of which have the same mass. We used the 
temperature and density histories of the tracer particles to calculate the nucleosynthetic yields in a 
post-processing step using libnucnet, and to 
provide additional diagnostics for the complex flows. Each simulation contained approximately 
$10^{4}$ tracer particles. 

\section{Results}
\label{Results}
This section is largely composed of two parts: the detailed analysis of the central-detonation models 
in $\S \ref{Detonation Structure for Non-Steady Stars}$, $\S  \ref{The Effect of Expansion Time Scales 
on Nucleosynthesis}$, and $\S \ref{An Empirical Relation}$, and the exploration of the off-center 
detonation models in $\S \ref{Off-Center Detonation}$.   

\subsection{Detonation Structure for Non-Steady Stars}
\label{Detonation Structure for Non-Steady Stars}
In this section, we discuss the properties of a one-dimensional spherical detonation as it 
propagates through a WD that is either expanding or contracting. In what follows, we distinguish 
among three types of hydrodynamic motion that interact nontrivially at different phases of the 
evolution of the explosion: the expanding or contracting flow belonging to the initial pre-expanded 
model; the expanding rarefaction that trails the detonation wave; and the bulk expansion experienced 
by the stellar material downstream of the detonation wave due to the energy release by the wave. We 
show that the combined effect of the 
first two flows is responsible for creating the specific cooling conditions behind the shock that 
determine the time available for nuclear burning, and hence the nucleosynthetic yields.

\subsubsection{Hydrodynamics of a Detonation Wave}
\label{Hydrodynamics of a Detonation Wave}
\begin{figure*}[h!]
\centering
\includegraphics[width=8cm, scale=1.0, angle=0]{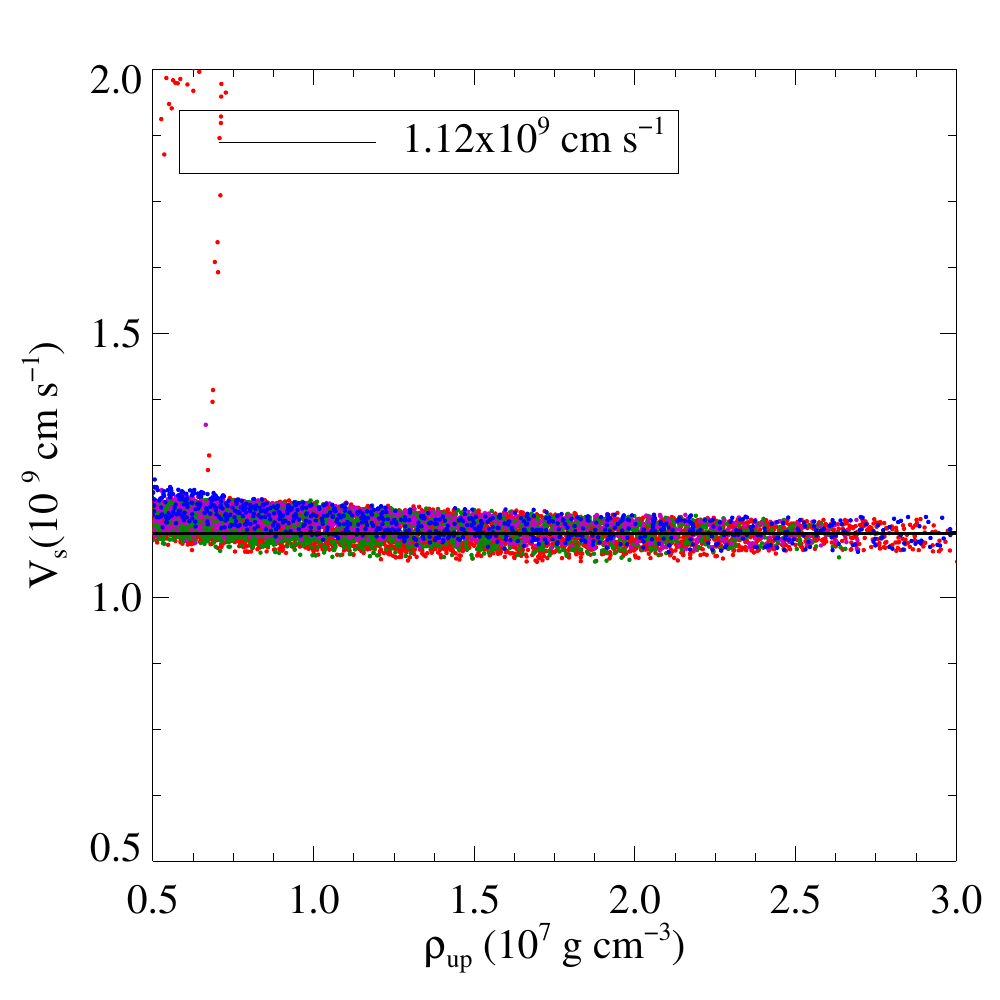}
\caption{The speed of the shock wave in the frame of the upstream fuels as a function of upstream 
density for four different models: HCV (blue), LCV
(magenta), HEV (red), and LEV (green). }
\label{sh_str}
\end{figure*}

The initial high-temperature disturbance in the center of the WD forms a shock wave that 
quickly transitions to a detonation whose wave speed is roughly the Chapman-Jouguet (CJ) speed 
$\sim \ 10^{9} \ cm/s$. The CJ speed corresponds to the smallest possible detonation speed, and 
many detonations fall into this category \citep{Sei09}. In our study, the detonation wave speed in 
the frame of the upstream fuel remains almost constant (varying by only $\pm 5\%$) as it propagates 
through 
the entire star, and this behavior is true in all models (see Figure~\ref{sh_str}). This implies that the 
shock strength is independent of the upstream velocities and is insensitive to the upstream density 
of the fuel.  

The propagation of the detonation wave is powered by the exothermic nuclear reactions that occur 
immediately behind it. The detonation wave
propagates outward from the center of the WD, consuming the carbon and oxygen in the stellar core. 
Carbon burning occurs almost instantly in the high-density regions close to the 
center of the WD, while its burning length scale becomes comparable to the maximum resolution of 
the simulation (500m) at densities lower than $\sim \ 1.0\times 10^{6} \ g~cm^{-3} $. The length scale 
for Oxygen burning becomes resolved at densities $< \ \sim \ 5\times 10^{6}$\ g~cm$^{-3}$, 
while the length scale for Silicon burning is larger than the maximum resolution at all densities 
\citep{Dom11}.  As the detonation wave approaches the WD surface, it is still burning almost all of the 
carbon but leaves unburned Oxygen 
in the low-density outer layers of the star. Not enough energy is released by nuclear burning behind 
the front in these low-density outers to continue to power the detonation wave; as a result, 
it transitions to a shock wave and propagates out of the star.   

\subsubsection{Rarefaction and Bulk Expansion}
\label{Rarefaction and Bulk Expansion}
\begin{figure*}[h!]
\centering
\includegraphics[width=16cm, angle=0]{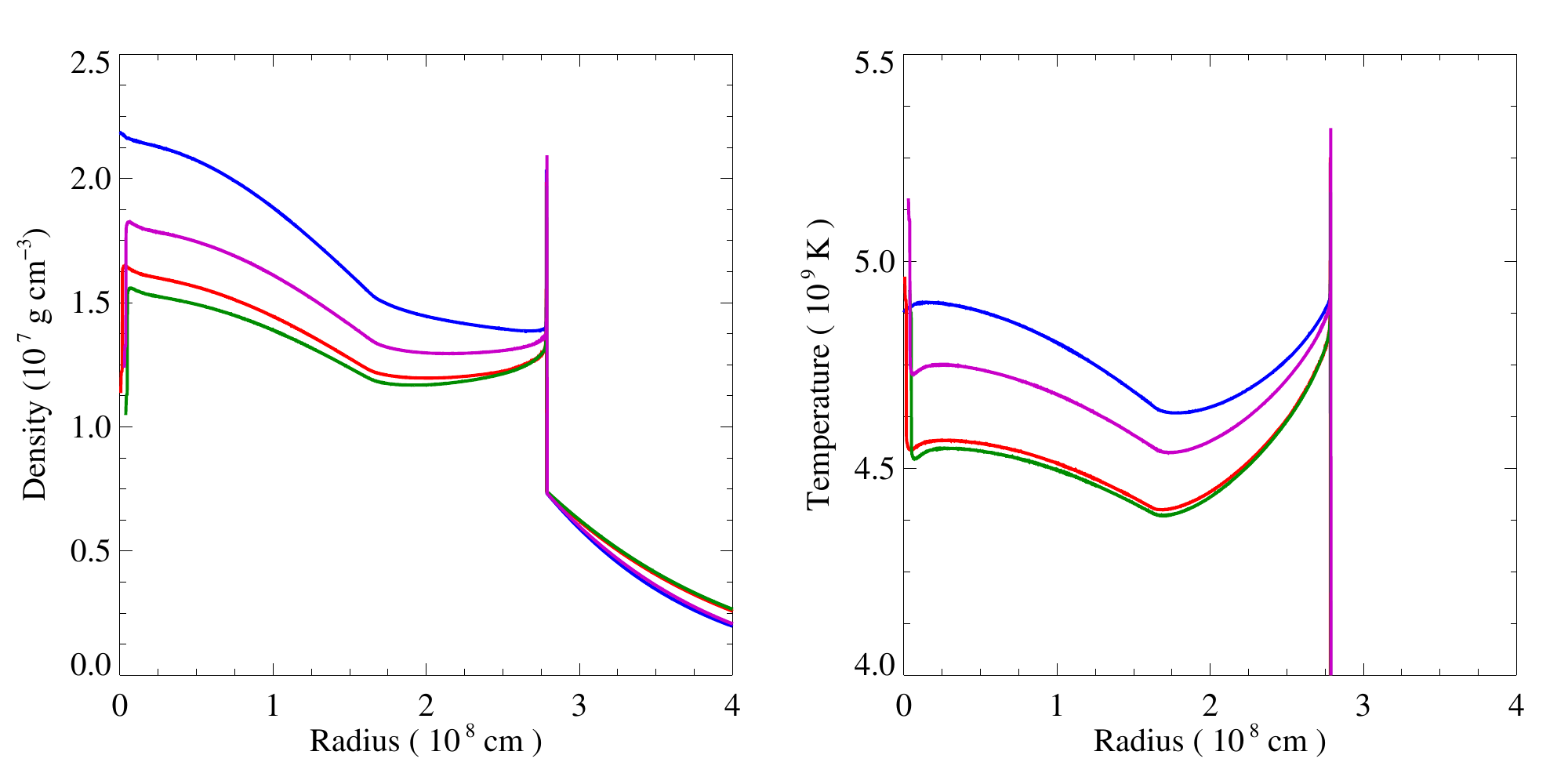}
\caption{Stellar density profiles (left panel) and temperature profiles (right panel) for 
four different models: HCV (blue), LCV 
(magenta), HEV (red), and LEV (green). The profiles are aligned such that their upstream densities 
overlap. The sharp discontinuity in both density and temperature profiles represents the detonation 
wave, which is traveling outward in radius. The structure of the thermodynamic properties behind the 
detonation wave front differs for the different models. }
\label{prof_d_t}
\end{figure*}

The temperature and density of the stellar material increase sharply due to the detonation wave; 
immediately afterward, the shocked material experiences a strong rarefaction. As an example, we 
show the behavior of four models at an upstream density of $7.5\times 10^{6} \ g\ cm^{-3}$. Figure~
\ref{prof_d_t} shows the density and temperature structure of the four models. Since the different 
models have different 
initial velocity profiles, the time it takes for the detonation wave to reach approximately the same upstream 
density differs. Therefore, we aligned the profiles so that their 
upstream densities overlap.

For the above upstream density, the matter in all models enter the detonation wave and is strongly 
compressed and heated to similar post-shock densities and temperatures ($2.1 \times 10^{7} \ g\ 
cm^{-3}$ and $5.3 \times 10^{9} \ K$). However, the strength of 
the rarefaction wave behind the front differs greatly, depending on the model. 
In general, the speed of the rarefaction wave behind the front is higher in expanding cores than in 
contracting ones. This trend is consistent with the thermal history of the tracer particles, which we will 
discuss in $\S \ref{The Effect of Expansion Time Scales on Nucleosynthesis}$. 

In view of the fact that
 the strength of the detonation wave is similar in all models, as illustrated in 
Figure~\ref{sh_str}, the difference in the expansion rates behind the detonation wave shows that the 
physics in the rarefaction zone is not dominated by the structure of the detonation wave; rather, 
it is a result of the superposition of the flow due to the detonation wave and that due to the pre-existing 
velocity profile in the star. This is the reason the nuclear yields are different in the different models;
 had the detonation structure dominated the preexisting flow, all of the models would have had similar 
 yields. Besides the rarefaction wave, there is another expanding phase due to the bulk expansion of
  the white dwarf.

\begin{figure*}[h!]
\centering
\includegraphics[width=16cm,  angle=90]{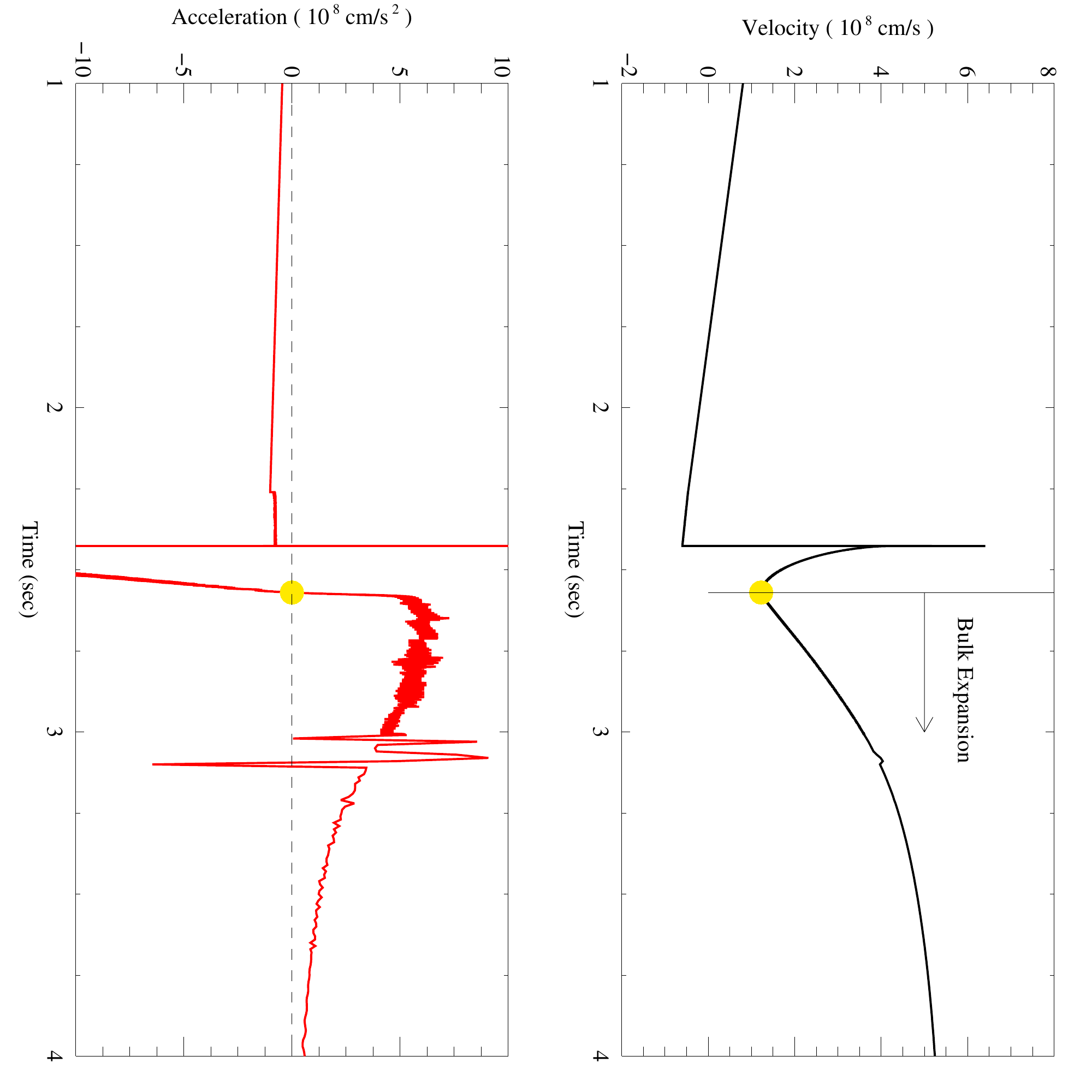}
\caption{Velocity and acceleration of a typical tracer particle as a function of time. Shown are the 
velocity (black curve) and the total acceleration (red curve) of the fluid as a function of time as given by
a typical tracer particle. At time t = 2.57 sec (indicated by the yellow dot), 
the fluid experiences positive acceleration and its outward velocity begins to increase with time.}
\label{tracer}
\end{figure*}

\begin{figure*}[h!]
\centering
\includegraphics[width=16cm, angle=90]{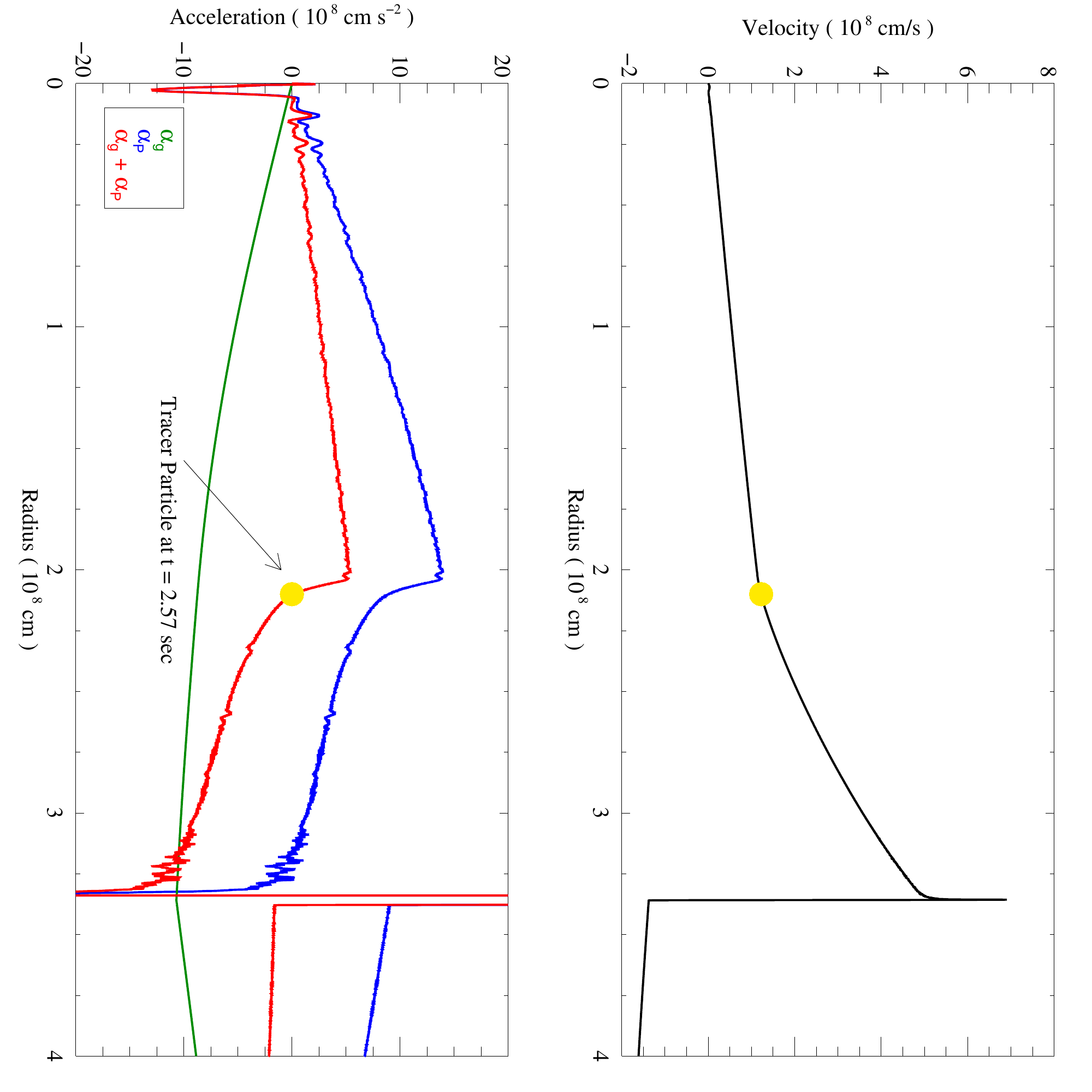}
\caption{Velocity and acceleration profile of the star. Shown are the velocity (black curve), the 
acceleration due 
to the pressure gradient (blue curve), the acceleration due to gravitational force (green curve), and the 
sum of the two accelerations (red curve). The yellow dot marks the location at time = 2.57 sec after the 
onset of the detonation of a tracer particle that was originally at a radius of $4.5 \times 10^{7} cm$ at $t$ = 0 sec.}
\label{hydro}
\end{figure*}

 Figures~\ref{tracer} and~\ref{hydro} 
show Lagrangian and Eulerian views of the typical flow properties behind the outwardly moving 
detonation wave for the \textbf{LCV} model. The yellow dot marks the position of a tracer 
particle in time in Figure~\ref{tracer} and in space in Figure~\ref{hydro}. The time histories of the tracer 
particle in velocity 
and acceleration are shown in Figure~\ref{tracer}. The tracer particle which is originally located at a radius of $4.5 \times 10^{7} cm$ at t = 0 sec travels towards the center of the WD while the detonation wave front propagates through the contracting star. As the tracer particle enters the detonation wave, it 
momentarily experiences a strong acceleration outward due to compression in the detonation wave. It is then accelerated toward the center (see the bottom panel of Figure~\ref{tracer}), as the pressure 
decays due to the expansion of the wave front. At t = 2.57 sec, the tracer particle experiences outward acceleration due to the bulk expansion of the star. The turnover in the acceleration can be seen 
in the time history of the tracer particle in velocity. This behavior is explained by 
Figure~\ref{hydro}, which shows the velocity profile and the acceleration  ($\alpha$) profile of 
the star t = 2.57 sec when the propagating 
detonation wave is $\sim 3.4 \times 10^8 cm$ from the center of the star. 
The pressure soon dominates over the inward gravitational force as the 
matter is heated by the energy that is released in the nuclear burning, leading to bulk 
expansion of the star (the location of the tracer particle is shown by the yellow dot).    

\subsection{The Effect of Expansion Time Scales on Nucleosynthesis}
\label{The Effect of Expansion Time Scales on Nucleosynthesis}
We calculated the final nucleosynthetic yields for each tracer particle by integrating the reaction 
network in libnucnet, starting when the tracer particle experienced the peak temperature $T_{peak}$ 
produced by the detonation wave. We continued the 
integration until the temperature decreased to $1.0 \times 10^{9} \ K$, below which 
the nuclear abundances do not change significantly except by beta decay. 
We found that, in all of the models, freeze-out to $^{56}$Ni occurs during the strong rarefaction 
phase and before the bulk expansion phase during which the pressure force drives the incinerated matter into 
expansion.  

\subsubsection{Nuclear Abundances}

\begin{figure*}[h!]
\centering
\includegraphics[width=16cm, angle=0]{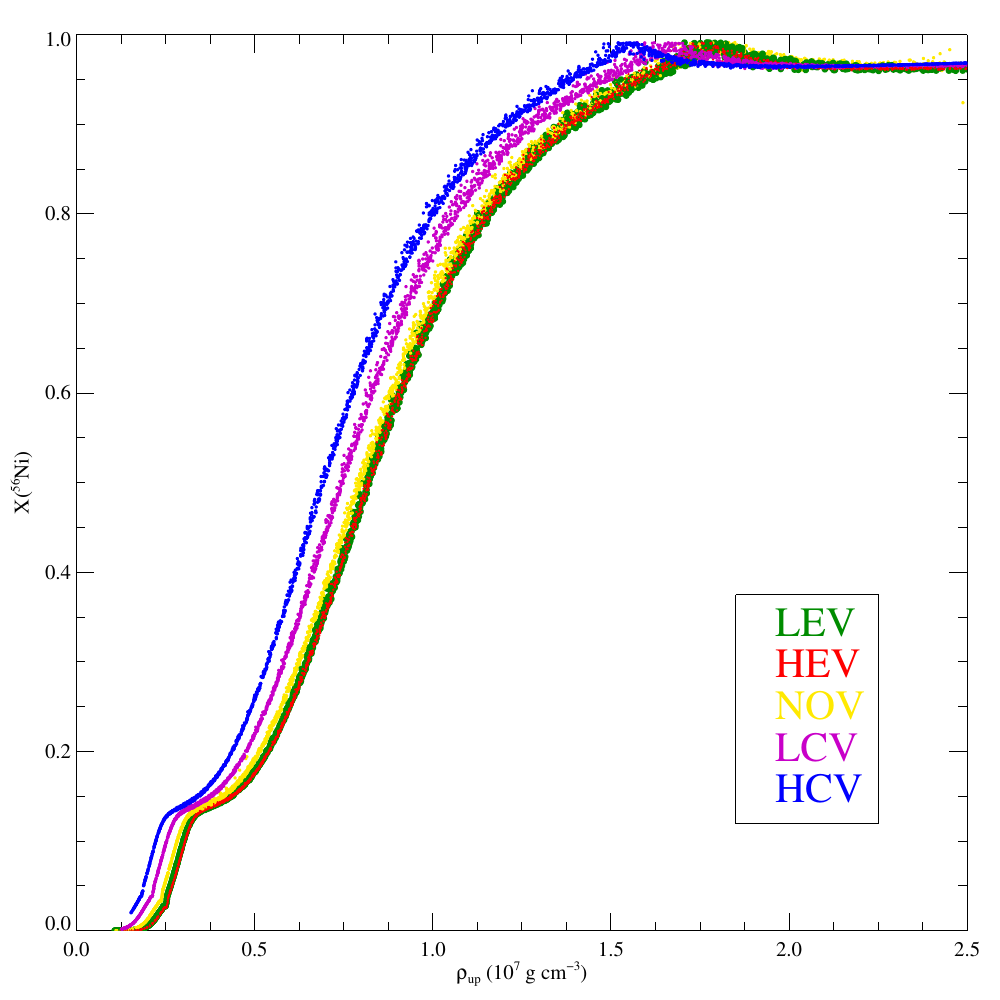}
\caption{Final mass fraction of $^{56}$Ni as a function of upstream density derived from tracer 
particles. }
\label{ni56_upD}
\end{figure*}

At the center of all of the models, the detonation wave converts the material 
chiefly into $^{56}$Ni by complete silicon burning. As the wave propagates through the 
low-density, outer layers, the nucleosynthesis is characterized by carbon and oxygen 
burning, then only carbon 
burning, and finally the burning ceases. Thus, the amount of $^{56}$Ni is, in general, a function of 
the initial density of the fuel, and increases with the fuel density. Figure~\ref{ni56_upD} shows the final 
mass fraction of $^{56}$Ni as a function of the upstream density $\rho_{up}$ of fuel, where 
``upstream'' denotes the 
undisturbed matter entering the detonation wave. While there is a tight relation between the final
abundance of $^{56}$Ni and $\rho_{up}$ in each simulation, we find that the final abundance of 
$^{56}$Ni for a given 
$\rho_{up}$ varies across the five simulations, with the $\mathbf{HEV}$ model producing the least 
$^{56}$Ni and the $\mathbf{HCV}$ model producing the most.

The differences in the final abundances of $^{56}$Ni across the five models are due to the differences 
in the bulk expansion time scales, as we demonstrate in the next section. 

\subsubsection{Time Evolution of Thermodynamic Properties}

In general,  the detailed behavior of the density and temperature is coupled to the 
nuclear burning and hydrodynamics. In the case of a simple one-dimensional detonation model, 
the detonation wave propagates outward from the point of ignition by heating  
the upstream fuel to a temperature above the ignition temperature. Once nuclear burning begins, the
energy released by it influences the hydrodynamic behavior of the ash, and 
the subsequent hydrodynamic behavior controls the time evolution of temperature and density of the ash. 
Because nuclear reactions are highly temperature-sensitive, the interplay of the 
time scales between burning and hydrodynamic expansion determines the final nucleosynthetic 
yields. 

\begin{figure*}[h!]
\centering
\includegraphics[width=16cm, angle=0]{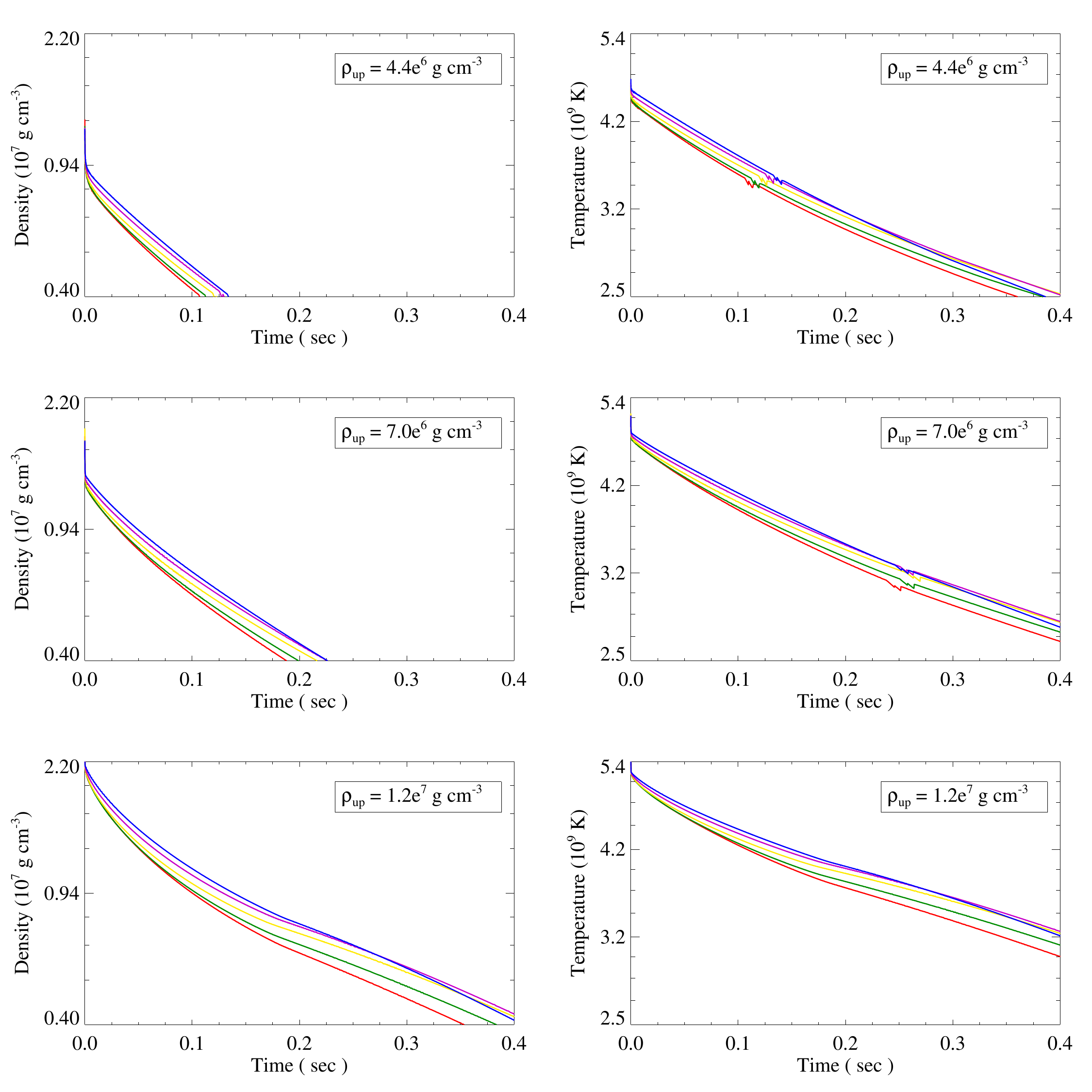}
\caption{Density and temperature histories of tracer particles whose upstream 
pre-detonation wave densities are 
$4.4 \times 10^{6} g\ cm^{-3}$, $7.0\times 10^{6} g\ cm^{-3}$, and $1.2 \times 10^{7} g\ cm^{-3}$ 
respectively from top to bottom. Shown are the results for five different models: $\mathbf{HCV}$ 
(blue), $\mathbf{LCV}$ (magenta), $\mathbf{NOV}$ (yellow), $\mathbf{LEV}$ (green), and $
\mathbf{HEV}$ (red).}
\label{time_his}
\end{figure*}

Figure~\ref{time_his} shows for all five models the density and temperature histories of tracer particles 
whose upstream densities are $4.4 \times \ 10^{6} \ g\ cm^{-3}$, $7.0 \times 10^{6} \ g\ cm^{-3}$, and 
$1.2 \times 10^{7} \ g\ cm^{-3}$, which spans the density range for which the final abundance of 
$^{56}$Ni 
lies between 0.1 and 1.0. The tracer particle histories were aligned such that their 
peak temperatures begin at time = 0.0 second.  

The peak density and the peak temperature are the same to within 3\% across the five models, 
but decrease on different time scales. We find that the flows with higher incoming 
speeds relative to the detonation wave (simulations: $\mathbf{HCV}$ and $\mathbf{LCV}$) expand 
more slowly compared to those with lower incoming speeds (simulations: $\mathbf{HEV}$ and $
\mathbf{LEV}$). Most burning to $^{56}$Ni occurs within $\sim$ 0.4 seconds after the temperature 
reaches its peak.

\subsubsection{The Effect of Expansion Time Scale on the Nuclear Yields}

Within the stellar material that has a final $^{56}$Ni mass fraction above 0.1, the peak
temperature is high enough to ensure silicon burning. The final $^{56}$Ni yield depends on the 
peak temperature and the expansion time scale. Freeze out occurs for a nuclide when the 
temperature drops low enough that reactions become too slow to alter its abundance. 
Because this condition occurs at different temperatures for different nuclides, the final abundances 
depend on the rate at which the material cools.    

\begin{figure*}[h!]
\centering
\includegraphics[width=8cm,angle=0]{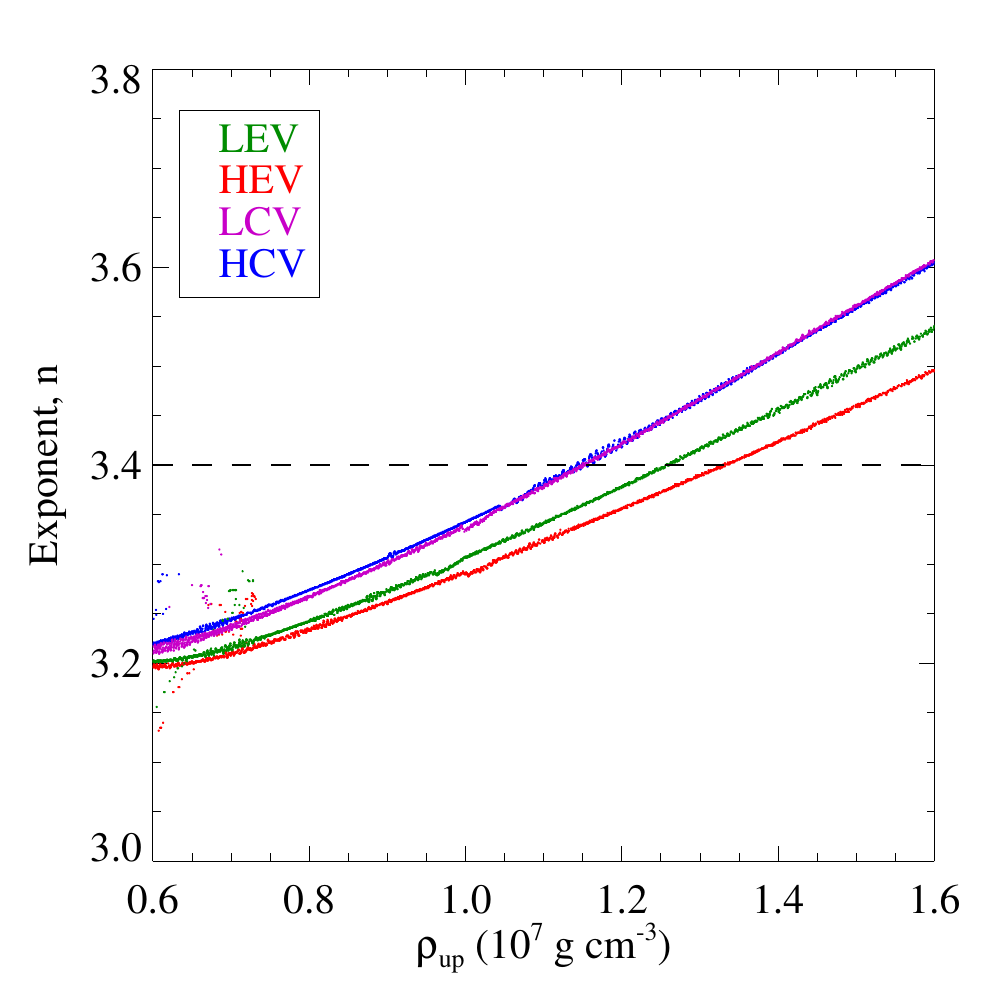}
\caption{Values of the exponent $n$ in equation (6) for the density, which were estimated from fits to a 
set of approximately 
7000 tracer particle time histories over the range in upstream density that yields a final mass fraction 
of $^{56}$Ni
between 0.1 and 1.0. The central value $n$ = 3.4 was adopted for calculations that investigated   
the sensitivity of the final yields to variations in the expansion time scale $\tau$. }
\label{exponent}
\end{figure*}

The thermodynamic trajectories  of the tracer particles (i.e. individual Lagrangian mass elements) 
processed by the detonation wave are well characterized by an exponential temperature dependence 
\citep{Arn71, Woo73, THW00, Mea09}  
\begin{equation}
T(t) = T_{0}\; \mathrm{exp}^{-t/\tau},
\end{equation}
where $T_{0}$ is the initial temperature at which the nuclear burning begins and $\tau$ is the time 
scale for the temperature to decrease to 1/e of its initial value. This time scale characterizes the 
expansion of the fluid after it is compressed by the leading shock. 
The density is related to the temperature by the fitted formula 
\begin{equation}
\rho(t) = \rho_{0}\; \left [\mathrm{exp}^{-t/\tau} \right ]^n,
\end{equation}
where n is a function of upstream density and varies between 3.2 and 3.6 over the range of densities 
for which the final $^{56}$Ni mass fraction lies above 0.1, as shown in Figure~\ref{exponent}. In this 
work, we adopt a central value of $n = 3.4$.  
 
\begin{figure*}[h!]
\centering
\includegraphics[width=8cm,angle=90]{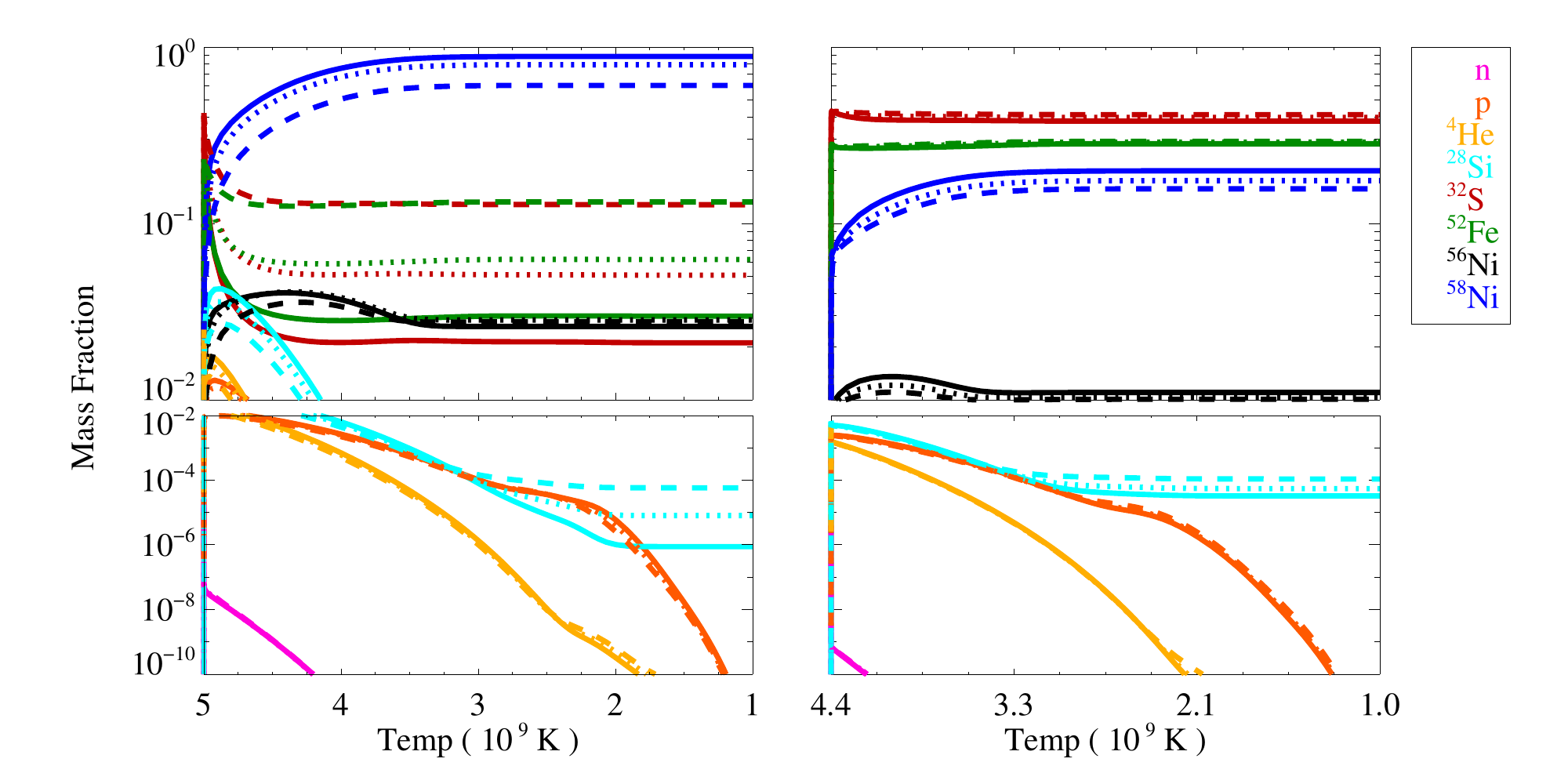}
\caption{Time evolution of selected light elements, intermediate mass elements, and iron group 
isotopes during the expansion, which is parameterized by temperature. Left panel: initial peak 
temperature of $5.0 \times 10^{9}$\ K and density of $1.2\times 10^{7}$\ g\ cm$^{-3}$.  Right panel: Initial peak 
temperature of $4.4 \times 10^{9}$ K and density of $9.4\times 10^{6}$\ g\ cm$^{-3}$. For each 
species, the abundances have been calculated with a nuclear network (see $\S\ref{Results}$) using an 
exponential temperature evolution for three different expansion time scales: 0.1 s (dashed), 0.3 s 
(dotted), and 0.6 s (solid). These figures illustrate the dependence of the nuclear 
abundances on the expansion time scale, especially for the heavy nuclei.}
\label{tau_dep}
\end{figure*}

In order to study the sensitivity of the final $^{56}$Ni abundance to variations in $\tau$, we 
used libnucnet to perform nucleosynthesis using thermal profiles that correspond to an 
exponential expansion. We chose peak density and temperature values that are characteristic of 
our tracer particle data. The initial composition consisted of equal masses of C and O, and the 
expansion timescale $\tau$ varied between 0.1 and 0.6, reflecting the estimated range seen in the 
tracer particle time histories in our models. The time evolution of selected isotopes is shown in 
Figure~\ref{tau_dep} for different expansion timescales and two different initial thermal conditions: an 
initial peak temperature of $5.0 \times
10^{9}$\ K and a density of $1.2\times 10^{7}$\ g\ cm$^{-3}$ (left panel), and an initial peak 
temperature of $4.4 \times 10^{9}$ K and density of 
$9.4\times 10^{6}$\ g\ cm$^{-3}$ (right panel).   

Stellar material with a similar initial density and temperature ends up with different abundances of $^{56}$Ni, 
depending on the history of the hydrodynamic expansion during the nuclear burning. Even in the case 
of stellar material that reaches the same peak temperature, the thermal profile with the longer 
(0.6 sec) expansion timescale produces more $^{56}$Ni while burning more $^{28}$Si than does 
with the shorter (0.1 sec) expansion timescale \citep{Cha12}.

\begin{figure*}[h!]
\centering
\includegraphics[width=8cm,angle=0]{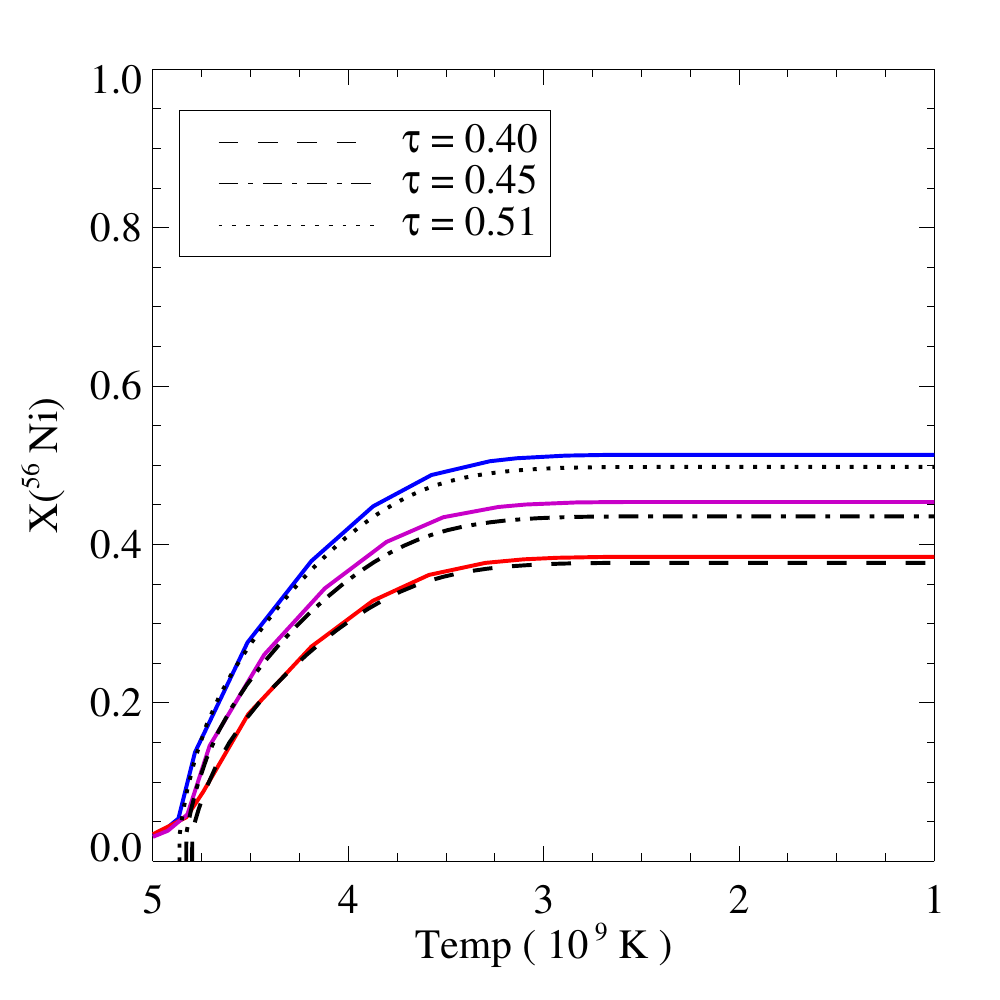}
\caption{Time evolution of $^{56}$Ni parameterized by the temperature. The solid lines represent 
the final $^{56}$Ni abundances derived from the density and temperature histories of tracer particles 
whose upstream densities 
are all $\rho = 7.0 \times 10^6\ g\ cm^{-3}$ for three different models: $\mathbf{HCV}$ (blue),
$\mathbf{LCV}$ (magenta), and $\mathbf{HEV}$ (red). The final $^{56}$Ni abundances derived from 
fitting equation (6) to tracer particle data are 
calculated for expansion time scales of 0.40 (dashed), 0.45 (dot-dashed), and 0.51 (dotted).  }
\label{tau_nuc}
\end{figure*}

To demonstrate this point, we compare the nucleosynthetic yield for tracer particles in different models 
with that from the exponential thermal profiles in (5) and (6). For a given upstream 
density of $\rho = 7.0 \times 10^6\ g\ cm^{-3}$, the temperature and density history of a tracer particle 
was analyzed and its expansion time scale estimated in three models: 0.4s in $\mathbf{HEV}$, 
0.45s in $\mathbf{LCV}$, and 0.51s in $\mathbf{HCV}$. Nucleosynthesis was 
performed using both the temperature history of the tracer particles and the exponential temperature 
profiles. We compare the results in Figure~\ref{tau_nuc}. 
The tracer particle from the model $\mathbf{HCV}$ model (shown in blue) has the longest expansion 
time scale, while the one from the $\mathbf{HEV}$ model (shown in red) has the shortest. Importantly, 
the nucleosynthesis produced by each time history is consistent with that produced by an exponential 
temperature profile with a similar  time scale. We conclude that the final $^{56}$Ni abundance is 
sensitive to the expansion time scale, and that the expansion time scale experienced by the tracer 
particles depends on the model.  It does so because the models differ in the velocity of the upstream 
stellar material relative to the detonation wave in the laboratory frame.   

\subsection{Identification of a Physical Parameter That Can Be Used to Estimate Nucleosynthetic Yields}
\label{An Empirical Relation}

In $\S\ref{The Effect of Expansion Time Scales on Nucleosynthesis}$, we showed that, even for similar
fuel densities upstream of the detonation wave, the final nucleosynthetic yields vary depending on the 
thermal expansion history behind the detonation wave. Since the expansion time scale depends on 
the incoming velocity, we have demonstrated that the upstream velocity affects the final 
nucleosynthetic yields. In this section, we identify a physical parameter that 
is correlated with the expansion time scale, and that can therefore be used to estimate the 
final nucleosynthetic yields.

\begin{figure*}[h!]
\centering
\includegraphics[width=8cm, angle=90]{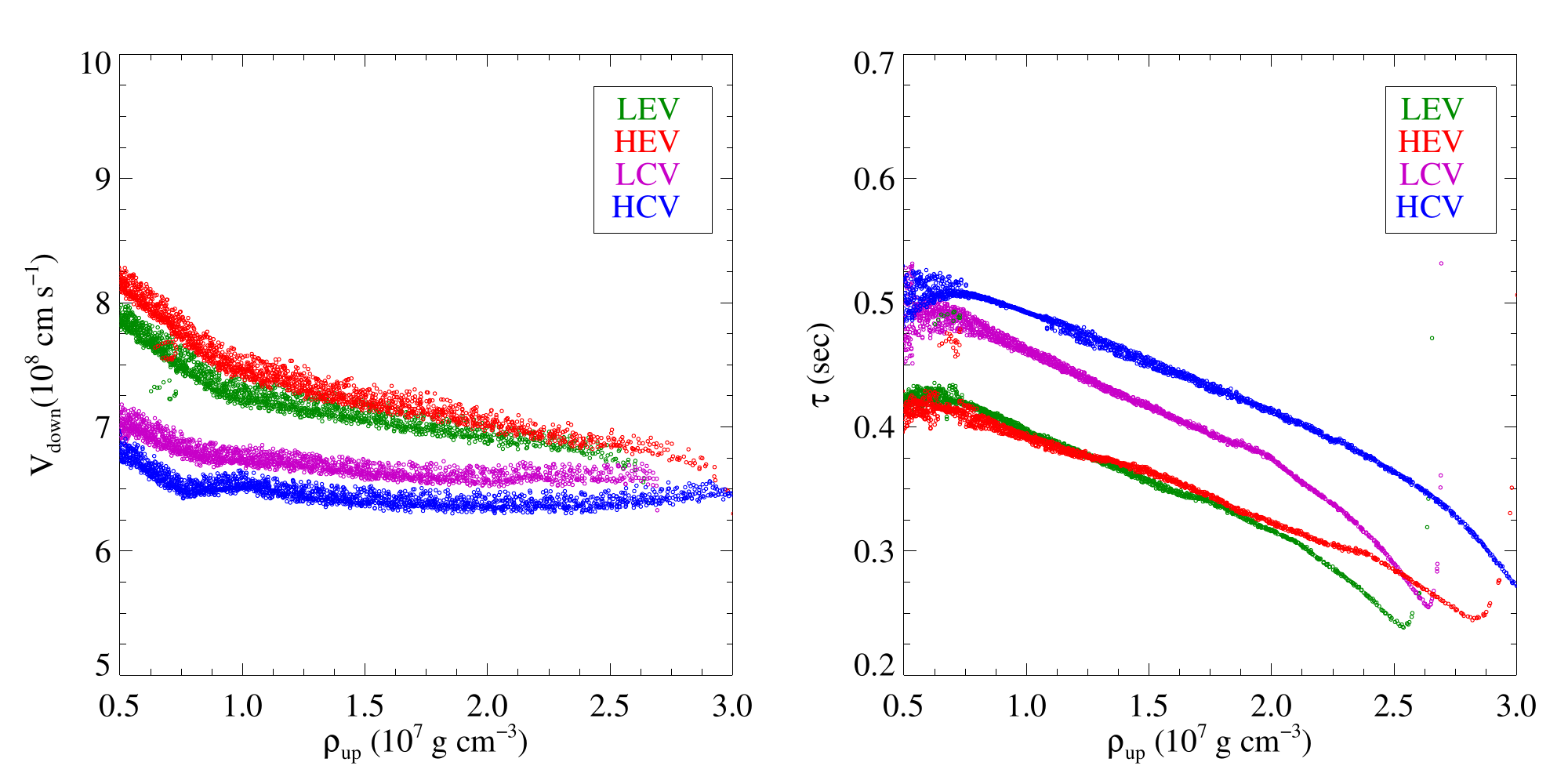}
\caption{Left panel: Downstream flow velocities derived from tracer particles as a function of 
upstream density. Right panel:  
Expansion time scales of the flow downstream of the detonation wave front as a function of upstream 
density. Shown are the results four different models: 
HCV (blue), LCV (magenta), HEV (red), and LEV (green).   }
\label{etau_peakv}
\end{figure*}

\begin{figure*}[h!]
\centering
\includegraphics[width=8cm, angle=0]{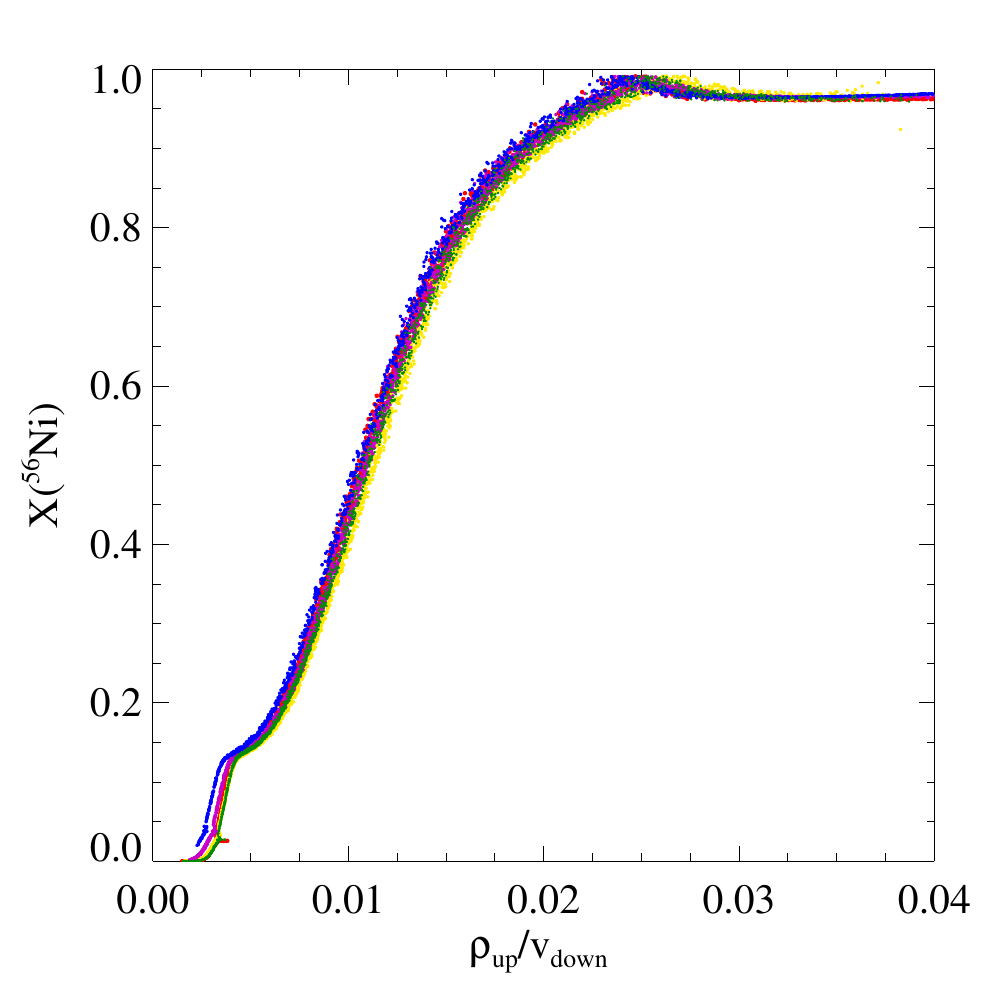}
\caption{Mass fraction of $^{56}$Ni as a function of the ratio of mass density immediately upstream 
of  and the velocity immediately downstream of the detonation wave front. Shown are the results for
five different models: HCV (blue), LCV (magenta), NOV (yellow), LEV (green), and HEV (red).   }
\label{scale}
\end{figure*}

The detonation wave connects two different flows: upstream and 
downstream. Here we define upstream and downstream as the physical states immediately before 
and after the detonation wave. For the same $\rho_{up}$, 
we found that the flow with a higher downstream velocity ($v_{down}$)
expands faster than the flow with a low downstream velocity. This is illustrated in 
Figure~\ref{etau_peakv}, and indicates that the downstream velocity is a second important parameter 
affecting the thermal expansion behind the shock front and therefore the nucleosynthetic yield.  In 
other words, flows with a higher downstream velocity yield lower $^{56}$Ni abundances. This 
suggests that a physical parameter formed by the ratio of $\rho_{up}$ and $v_{down}$, i.e. $\rho_{up}/
v_{down}$, might be an even better estimator of the final abundance of $^{56}$Ni than $\rho_{up}$ 
alone.  Figure~\ref{scale} shows that this is indeed the case: the physical parameter formed by the 
ratio of upstream density and downstream velocity is able to predict the final abundance of $^{56}$Ni to within 10$\%$.    

\subsubsection{Intermediate Mass Elements}
\begin{figure*}[h!]
\centering
\includegraphics[width=16cm, angle=0]{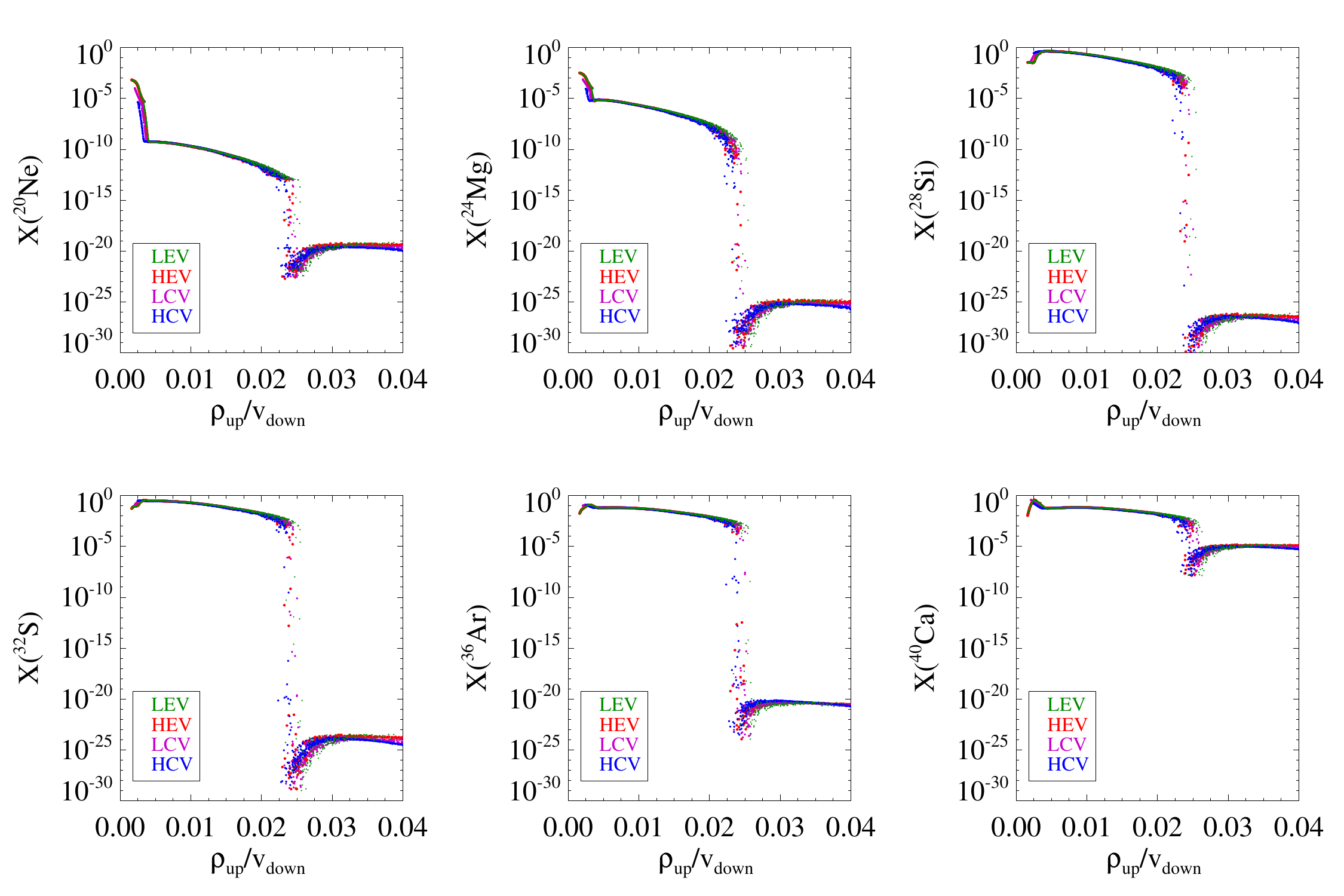}
\caption{Mass fraction of the intermediate-mass elements:$^{20}$Ne, $^{24}$Mg, 
$^{28}$Si, $^{32}$S, $^{36}$Ar, and $^{40}$Ca as a function of the ratio of $\rho_{up}/v_{down}$. 
Shown are the results for four different models: HCV (blue), LCV (magenta), LEV (green), and HEV 
(red).   }
\label{int_scale}
\end{figure*}

The timescale to burn to $^{56}$Ni is much longer than those to burn to the primary intermediate mass 
elements such as $^{20}$Ne, $^{24}$Mg, $^{28}$Si, $^{32}$S, $^{36}$Ar, and $^{40}$Ca. We 
therefore hypothesize that the physical parameter $\rho_{up}/v_{down}$ should be a good predictor 
for the final abundances of intermediate mass elements as well. Figure~\ref{int_scale}, which shows 
the final mass fraction of the intermediate mass elements as a function of the ratio of $\rho_{up}/
v_{down}$, shows that it is. 

\begin{figure*}[h!]
\centering
\includegraphics[width=16cm, angle=0]{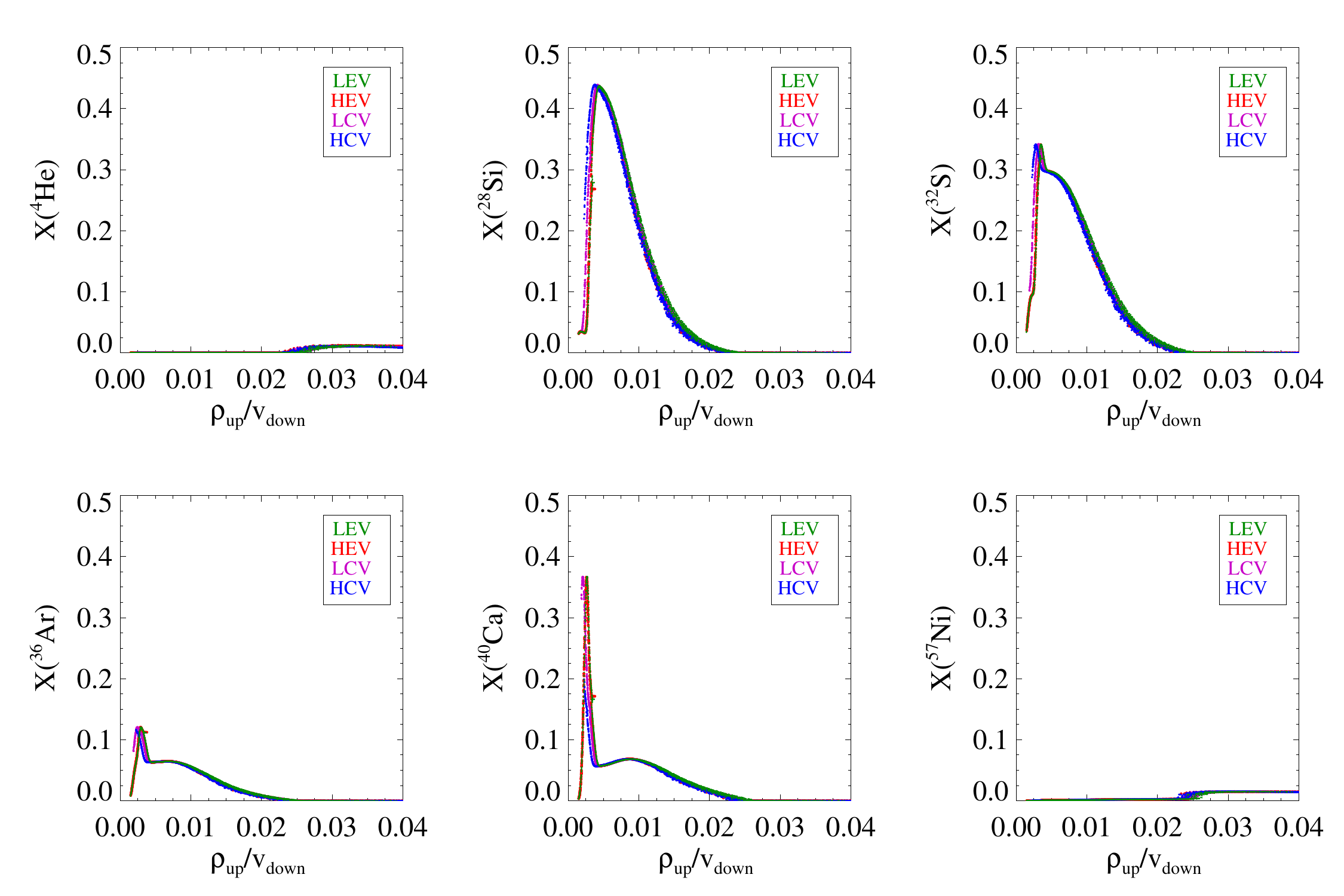}
\caption{Mass fraction of the dominant nuclei in mass fraction after $^{56}$Ni: $^{4}$He, $^{28}$Si, 
$^{32}$S, $^{36}$Ar, $^{40}$Ca, and $^{57}$Ni as a function of the ratio of $\rho_{up}/v_{down}$ in 
the region where the empirical relation holds. Shown are the results for four different models: HCV 
(blue), LCV (magenta), LEV (green), and HEV (red).    }
\label{scale_lin}
\end{figure*}

We also note two distinctive features in Figure~\ref{int_scale}. First, on the high-density branch (where 
$\rho_{up}/v_{down} > 0.025$), the 
intermediate mass elements form at the expense of $^{56}$Ni. This is most likely due to the
increased in entropy this mixture of elements represents. Second, on the low density branch (where $
\rho_{up}/v_{down} \sim 0.0$), 
the light $\alpha$-nuclei capture $^{4}$He and burn to heavier $\alpha$-nuclei, thus reducing the 
mass fractions of the lighter nuclei. 
The mass fractions of the most abundant elements in the $\rho_{up}/v_{down}$ range of interest are 
shown in the Figure~\ref{scale_lin}. While $^{56}$Ni is the dominant product in the high density 
region, at lower densities, relaxation to NSE is incomplete and the final product is a mixture of 
intermediate mass elements.

\subsection{Off-Center Detonation Models}
\label{Off-Center Detonation}
We have also investigated off-center detonation 
scenarios using one-dimensional models detonated at a finite radius. These models enable us to determine 
whether the empirical relation we found in $\S \ref{An Empirical Relation}$ also holds in off-center (but 
still spherically symmetric) detonation models. To investigate this question, we used the same four 
initial density and velocity profiles as before: $\mathbf{HCV}$, $\mathbf{LCV}$, $\mathbf{LEV}$, 
and $\mathbf{HEV}$. We initiated an off-centered detonation in each at a distance of $2.0 \times 
10^8$ cm from the stellar center.  

\begin{figure*}[h!]
\centering
\includegraphics[width=8cm, angle=0]{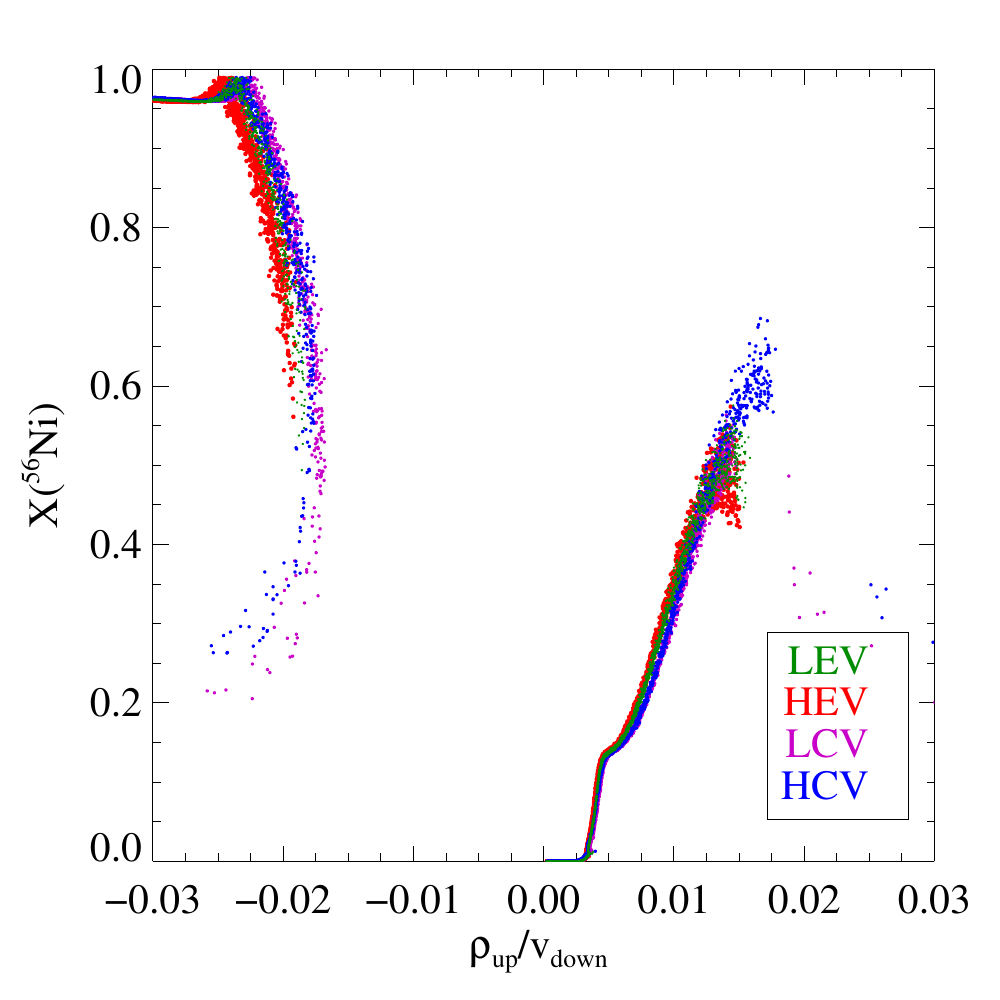}
\caption{Mass fraction of $^{56}$Ni as a function of the ratio of mass density immediately upstream 
of  and the velocity immediately downstream of the detonation wave front. Shown are the results for 
four different
off-center detonation models: $\mathbf{HCV}$ (blue), $\mathbf{LCV}$ (magenta), $
\mathbf{LEV}$ (green), and $\mathbf{HEV}$ (red).   }
\label{off_scale}
\end{figure*}

Figure~\ref{off_scale} shows the final mass fraction of $^{56}Ni$ 
as a function of the empirical physical parameter $\rho_{up}/v_{down}$ for the four off-center detonation 
models. In the off-center 
detonation models, detonation produces an ingoing detonation wave as well as an outgoing detonation wave.  The latter converges toward the center of the star, triggering a second explosion. In 
the region 
of the star where nuclear burning takes place as a result of the ingoing detonation 
wave, matter is driven inward by the detonation wave and the star is contracting. Consequently, v$_{down}$ is negative in this region, as is the ratio of 
$\rho_{up}$ and v$_{down}$. An empirical relation still holds, but it is  different for this contracting (high-density) branch
where $\rho_{up}/v_{down}$ $<0$ than for the expanding (low-density) branch. 

\begin{figure*}[h!]
\centering
\includegraphics[width=8cm, angle=90]{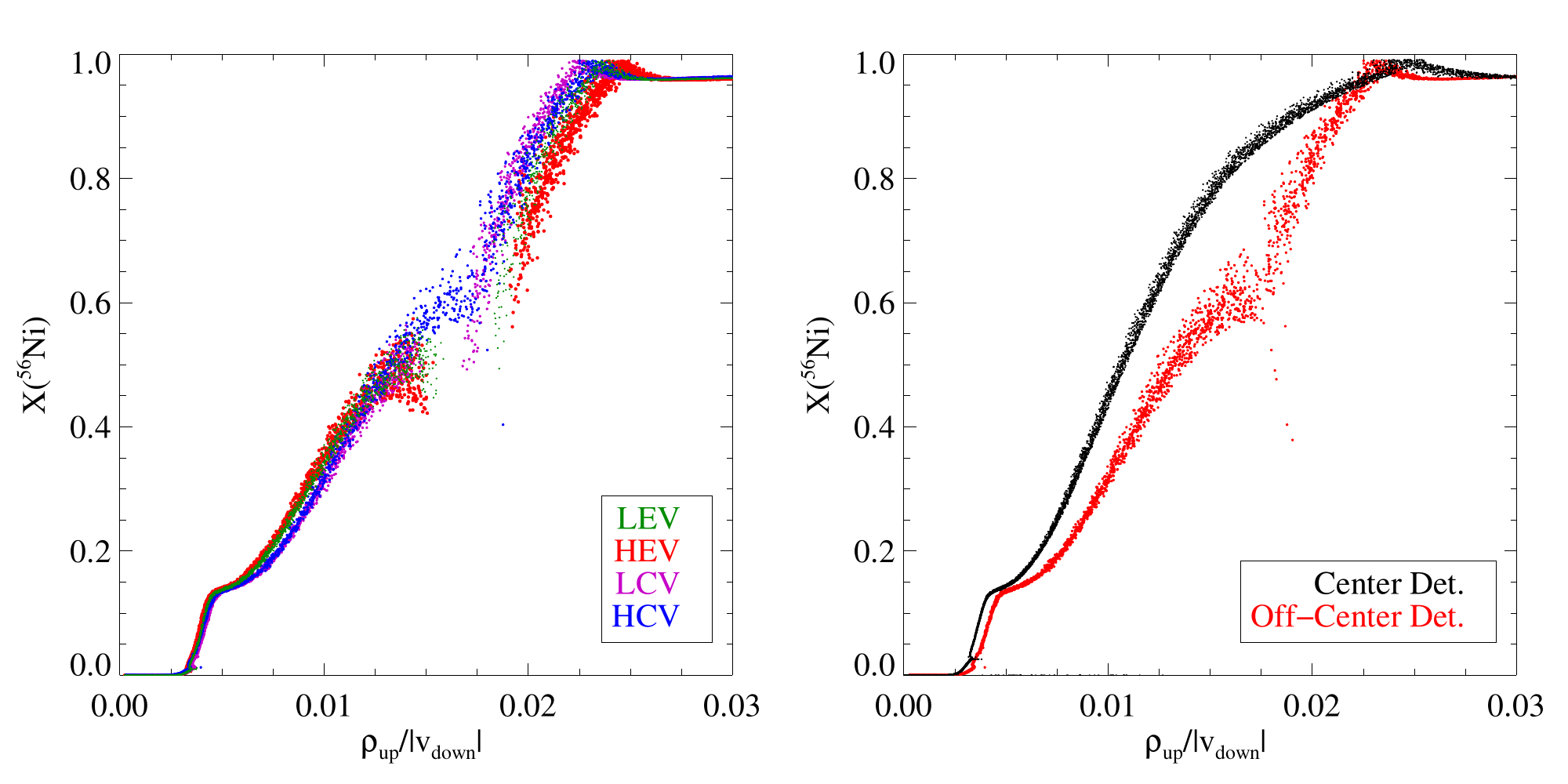}
\caption{Left panel: Mass fraction of $^{56}$Ni as a function of $\rho_{up}/|v_{down}|$. 
Shown are the results of four different off-center models: HCV (blue), LCV (magenta), LEV (green), 
and 
HEV (red). Right panel: the relation between $^{56}$Ni and $\rho_{up}/|v_{down}|$ for the off-center 
detonation model (red) is compared to the empirical relation in the central detonation model (black). 
Both models had an HCV initial profile.  }
\label{scale_comp}
\end{figure*}

In order to better compare the relation between the physical parameter $\rho_{up}/v_{down}$ and the final mass fraction of  
$^{56}$Ni for the two regions, we replace v$_{down}$ with its absolute value; the parameter then becomes $\rho_{up}/|v_{down}|$. The resulting relation is 
shown in the left panel of Figure~\ref{scale_comp}. The 
data in the vicinity of the detonation point are numerically noisy, and we have omitted them from the 
figure. The two branches join when plotted in this way.  Although 
the empirical relation is not as tight for the contracting (high-density) branch as it is for the expanding (low-density) branch, the 
physical variable $\rho_{up}/|v_{down}|$ is able to predict the final abundance of  $^{56}$Ni to 
within 15$\%$. This suggests that 
this physical parameter is a fairly robust predictor of the final nucleosynthetic yields. 

The right panel in the Figure~\ref{scale_comp} compares the empirical relation between $\rho_{up}/|
v_{down}|$ and the final abundance of  $^{56}$Ni for the center and off-center detonation models for 
the $\mathbf{HCV}$ initial stellar velocity profile. The results demonstrate again that the final nucleosynthetic 
yields cannot be characterized by the local values of density and velocity of the matter alone. Rather, 
they are affected by 
both the pre-existing velocity flow in the star and the post-detonation expansion of the star. 

\section{Conclusion}
\label{Conclusion}
The explosion mechanism of Type Ia SNe is still an active topic of research, and among the leading paradigms for the explosion mechanism are the delayed detonation models. While many different delayed detonation scenarios have been published, we direct our attention to investigating the key common feature of the explosion that involves the propagation of the detonation wave in a non-static velocity field of a white dwarf. Although the most satisfactory approach to this investigation would be to simulate a multidimensional delayed detonation model on scales of an exploding white dwarf, this is a numerically challenging task, requiring a considerable amount of data storage and computation time. Furthermore, our primary interest lies in the role played by the velocity profile settled during the previous deflagrative phase in the evolution of the detonation. Therefore, we create simplified one-dimensional models that test mainly the effects of the pre-detonation stellar internal velocity profile and the post-detonation velocity of expansion on the production of $\alpha$-particle nuclei, including $^{56}$Ni.   

We have studied the flow structure behind the traveling detonation wave front and observed two 
distinct expansion phases: rarefaction just behind the detonation wave and bulk expansion of the 
expanding white dwarf. The rarefaction timescale, or the 
expansion time scale studied in this work, was found to be not a function of detonation strength 
only, but is also influenced by pre-existing flow structure, depending on whether the initial 
configuration of the star is expanding or contracting. In our models, almost all the burning to $^{56}$Ni 
occurred only in rarefaction phase.

The relationship between the flow properties and the 
resulting $^{56}$Ni yields showed that the final yields of burned matter emerging from 
the detonation are highly sensitive to the expansion time scales of the flow behind the detonation, 
which is strongly conditioned by the pre-existing flow in the expanding or contracting progenitor 
star. The expansion time is longer for contracting stars and shorter for expanding stars, as the 
rarefaction proper to the detonation is combined with the pre-existing flow. With the greater 
expansion time scale, both temperature and density evolve relatively slowly, providing more time 
for $^{56}$Ni production.   
 
We have also found an empirical relationship between the ratio of the upstream density to the 
post-shock velocity and the final $^{56}$Ni yield for central detonations. It is not surprising to find that 
$^{56}$Ni yield increases with increasing upstream density of fuel. However, 
we also find that the final post-detonation velocity is another parameter that influences $^{56}$Ni
yield. The quantity $\rho_{up}$/$v_{down}$ is tightly related with $^{56}$Ni yield, and produces 
relatively small scatter ($< 10\%$) about the relationship; however, the picture is more complicated for 
off-centered detonations. Currently this relationship is limited to one-dimensional models and we caution that the above results are preliminary. In more realistic SNe Ia models, 
detonations in white dwarfs not only show a rich multilevel cellular structure \citep{Gam99,Tim00} but also travel through 
non-spherically symmetric stellar density distributions \citep{Jor08, Mea09}. In addition, some explosion scenarios involve 
detonations propagating through partially burnt material as well as unburnt fuel. Thus, further analysis in 
higher dimensions will need to be carried out. 

\newpage

\end{document}